\documentclass[12pt]{article}

\usepackage{epsfig}
\usepackage[psamsfonts]{amsfonts}
\usepackage{rotating}

\usepackage{graphicx}
\usepackage{amssymb}
\usepackage{slashbox}
\usepackage{color}

\newcommand{\qq}{q\bar q}
\newcommand{\Q}{{\cal Q}}
\newcommand{\QQA}{\Q _1 \bar \Q _1}
\newcommand{\QQB}{\Q _2 \bar \Q _2}
\newcommand{\QQi}{\Q _i \bar \Q _i}

\newcommand{\QQ}{\Q \bar \Q}

\newcommand{\jpc}{J^{PC}}
\newcommand{\slj}{^{2S+1}L_J}
\newcommand{\nslj}{n^{2S+1}L_J}

\newcommand{\8}{{\bf 8}}
\newcommand{\ten}{{\bf 10}}

\newcommand{\1}{{\bf 1}}
\newcommand{\3}{{\bf 3}}
\newcommand{\4}{{\bf 4}}
\newcommand{\6}{{\bf 6}}

\newcommand{\be}{\begin{equation}}
\newcommand{\ee}{\end{equation}}
\newcommand{\bea}{\begin{eqnarray}}
\newcommand{\eea}{\end{eqnarray}}
\newcommand{\no}{\noindent}
\newcommand{\non}{\nonumber}

\begin{document} 

\renewcommand{\baselinestretch}{1.3}

\normalsize
\begin{titlepage}

\begin{quote} \end{quote}
\vskip .1cm

\begin{center}

{\bf \Large N\normalsize O \Large R\normalsize ADIAL \Large E\normalsize XCITATIONS IN \Large L\normalsize OW \Large E\normalsize NERGY \Large QCD. I.

\vskip .4cm

D\normalsize IQUARKS AND \Large C\normalsize LASSIFICATION OF \Large M\normalsize ESONS}

\end{center}

\vskip .5cm
\large
\centerline{{\sc Tamar Friedmann$^*$$^\dagger$}}
\vskip .2cm
\normalsize

\centerline{\sl Massachusetts Institute of Technology, Cambridge, MA 02139, USA}

\vskip .2cm

\centerline{\sl University of Rochester, Rochester, NY 14623, USA}

\LARGE
\centerline{\it }
\normalsize

\renewcommand{\baselinestretch}{1.2}
\normalsize

\abstract{We propose a new schematic model for mesons in which the building blocks are quarks and flavor-antisymmetric diquarks. 
The outcome is a new classification of the entire meson spectrum 
into quark-antiquark and diquark-antidiquark states  
which does not give rise to a radial quantum number: 
all mesons which have so far been believed to be radially excited are orbitally excited diquark--antidiquark states; similarly, there are no radially excited baryons. Further, mesons that were previously viewed as "exotic" are 
no longer exotic as they are now naturally integrated into 
the classification as diquark-antidiquark states. 
 The classification also leads to the introduction of {\it isorons} (iso-hadrons), 
which are analogs of atomic isotopes, 
and  their {\it magic quantum numbers}, which are analogs of the magic numbers of the nuclear shell model. The magic quantum numbers of isorons match  the quantum numbers expected for low-lying glueballs in lattice QCD.
We observe that interquark forces in mesons behave substantially differently  from those in baryons: qualitatively, they are color--magnetic in mesons but color--electrostatic in baryons. 
We comment on potential models and the hydrogen atom.
The implications of our results for confinement, asymptotic freedom, and a new set of relations between two fundamental properties of hadrons -- their size and their energy -- are discussed in our companion paper \cite{companionII}.
}

\vfill 
\no $^*$E-mail: tamarf at mit.edu 

\no $^\dagger$Current E-mail: tamarf at pas.rochester.edu

\end{titlepage}



\renewcommand{\baselinestretch}{1.4}
\normalsize

\newpage


\section{Introduction}

It is well-known that making reliable predictions about low-energy QCD and hadrons is a great challenge, as perturbative methods of quantum field theory do not apply at low energies where the coupling constant is strong. The common approach has been to propose various dynamical models which are inspired by assumptions, ideas, and intuition borrowed from physical systems, such as atomic physics and non-relativistic quantum mechanics, which are not  QCD. 

In this paper we set out to study the hadron spectrum by employing purely QCD ingredients and invoking the role of diquarks in the mix. 

One well-established pillar of QCD is the quark model \cite{quarkmodel}, which has been the accepted framework for classifying the hadron spectrum. This is a schematic model for the mesons and baryons in which quarks are the building blocks for all the hadrons: mesons are bound states of a quark and an antiquark ($\qq$) and baryons are bound states of three quarks ($qqq$). 
In addition to quarks, 
bound configurations of two quarks, known as diquarks, may also be building blocks. 
The diquarks, explored at the beginning of the quark model in  the 1960's having been introduced already in Gell-Mann's paper \cite{quarkmodel} (for reviews see \cite{Anselmino:1992vg,Amsler:2004ps,Jaffe:2004ph}), have been revisited 
 following a surge of experimental and theoretical interest  in pentaquarks  ($qqqq\bar q$)\cite{Nakano:2003qx,Jaffe:2004ph,Jaffe:2003sg}.\footnote{Experiments eventually showed that the pentaquark $\Theta ^+$ does not exist \cite{Beringer:1900zz};  as Robert Jaffe said  (Harvard seminar, 2004), "pentaquarks  might come and go, but the diquarks are here to stay."} 
In particular, diquarks have been used as building blocks in a systematic classification of all known baryons \cite{Wilczek:2004im,Selem:2006nd}.
As to mesons, a few mesons have been 
viewed as having 
diquarks as constituents  -- to name just two examples, the light scalar mesons were interpreted as diquark-antidiquark states \cite{Jaffe:1976ig}, as were several charmed and hidden-charm mesons \cite{Rosenzweig:1975fq, CHARM}. But
 diquarks have never been employed systematically as building blocks for the classification of {\it all} known mesons. 

We undertake this task. Our purpose  is to find out whether the entire meson spectrum can be re--classified with the aid of diquarks, and whether we can learn anything new about QCD in the process.

In this spirit, we construct a 
new extended schematic model for mesons in which certain diquark configurations, selected for us by the  flavor structure of meson phenomenology, are building blocks for mesons in addition to and on equal footing with the quarks of the traditional quark model. These diquarks are the two flavor-antisymmetric ones.
One of the two coincides with the most well--known "good" diquark which is antisymmetric in all quantum numbers; the other has been previously unfairly neglected.

What follows is a reclassification of the meson spectrum into quark-antiquark and diquark-antidiquark states, with a reassignment of $L$ and $S$ quantum numbers to the mesons;  
mesons that were previously viewed as "exotics" are no longer exotic as they are now  naturally integrated into the classification as diquark-antidiquark states.

In the process, a new notion of {\it isorons} (iso-hadrons) 
emerges,
 along with their {\it magic $\jpc$ quantum numbers}. The isorons  are the natural analogs of isotopes or isotones in atomic or nuclear physics, 
and their magic $\jpc$ quantum numbers are analogous to the magic numbers of the nuclear shell model. In the nuclear shell model, it was spin-orbit couplings which was the magic behind the magic numbers. Here, it remains an open problem to understand what is behind the magic $\jpc$ of isorons. It is striking that the magic $\jpc$ of isorons match the quantum numbers predicted for low-lying glueballs by lattice QCD.

 Most significantly, we find that 
no radial quantum number appears
in the classification. 
In both the light and heavy quark sectors, 
mesons that have been believed to be radially excited quark--antiquark states are orbitally excited diquark--antidiquark states. 
The same is true for baryons: the baryons that have so far been considered to be radially excited 
appear to be an orbitally excited configuration of two diquarks and an antiquark.
All in all, 
the classification leads to the conclusion that 
there are no radial excitations in the hadron spectrum. 
In turn, this leads to inescapable, surprising, and significant implications 
regarding the dynamics of the strong force, confinement, and asymptotic freedom. In particular, 
they uncover 
 a new set of relations between two fundamental properties of hadrons: their size and their energy. These relations predict that hadrons shrink. They
are treated separately in our companion paper 
\cite{companionII} 

While our results may appear counterintuitive, they are completely consistent with the known properties of QCD, such as confinement and asymptotic freedom, and provide a novel explanation for the relation between them. We discuss this in our companion paper \cite{companionII}.

By now these results have experimental support: in \cite{pohl} (repeated in \cite{antognini}), the discovery of a shrunk size for the proton appeared, nine months after the original version \cite{arxiv} of this work was posted to arXiv.org, and in \cite{rhosize}, it is reported that the HERMES experiment found shrinkage of the $\rho$ meson; furthermore, charged bottomonium-like "exotic" states, the $Z_b(10610)$ and $Z_b(10650)$, were discovered by the BELLE collaboration \cite{Belle:2011aa}, which fit nicely in our classification tables as our predicted isovector made of a diquark and an antidiquark. 
Further experiments are suggested in the companion paper. 

In light of these experiments, we believe that the approach we took here is worthwhile and relevant, in spite of the fact that our results and conclusions are substantially different from those coming from other known models, such as potential models. 

This paper is organized as follows. In Section \ref{model}, we construct the new extended schematic model. We explain which diquark configurations constitute the building blocks for mesons, and we derive the color, flavor, and spin quantum numbers for mesons. In Section \ref{reclass}, we carry out the re-classification of mesons based on this model, and present the result in three tables: one for the light mesons, one for heavy mesons, and one for the isorons.  We then devote several subsections to a discussion of the resulting classification: in \ref{isoronsmagic}, we define isorons and their magic quantum numbers, and discuss their relations to quantum numbers of glueballs expected from lattice QCD; in \ref{norad}, we discuss the main result that  no radial quantum number arises in the classification and review the history of the radial quantum number and the difficulties in both theory and experiment that have surrounded this quantum number in the past; in  \ref{noexot}, we  discuss exotics.  In the rest of this section, we discuss various additional aspects of the classification, including predictions for new particles, inverted mass hierarchies, the binding energy of diquarks, and decays of diquark-antidiquark mesons.  In Section \ref{baryons} we turn to the baryon sector, showing that there, too, there are no radials.
In Section 5, we point out that the interquark forces appear qualitatively different in the meson and baryon sectors. Section 6 includes Regge trajectories of mesons. The Appendix includes a nonet by nonet discussion. 

\renewcommand{\baselinestretch}{1.4}
\normalsize

\section{Extended Schematic Model for Mesons}\label{model}

\subsection{A few good diquarks}\label{afew}

The first question we are faced with when constructing our model is which diquark configurations 
are 
building blocks for mesons, in addition to and on equal footing with the quarks of the quark model.

This question would be easy to answer if the interquark forces of low--energy QCD were known; if that were the case, we would know which diquark configurations are attractive and those would be  our building blocks. 
Since this is not the case, we instead derive the diquark building blocks from the following aspect of meson phenomenology: 
 in the light meson sector, where the flavor group is $SU(3)$, all observed meson multiplets are flavor nonets. Therefore, the diquarks must be those for which  diquark-antidiquark configurations would form only flavor nonets. 
 
 To figure out which diquarks satisfy this requirement, we first note that since light quarks are in the $\3 _f$ flavor representation, light diquarks  may form either flavor sextets or antitriplets:
\be \label{diquarkrep} \Q =  qq : \hskip .4cm \3 _f\otimes \3 _f = \6 _f\oplus \bar \3 _f ~ \hskip 2cm SU(3)_f,\ee
where $\Q$ stands for a diquark. The sextet $\6 _f$  is symmetric under flavor exchange of the two quarks, while the antitriplet $\bar \3 _f$ is antisymmetric under this exchange. Now, the only combination of a diquark and an antidiquark that forms exactly a flavor nonet and no larger multiplet is one in which the diquark is a flavor antitriplet and the antidiquark is a flavor triplet. This combination leads to the representations 
\be \label{insp}\QQ : \hskip .4cm \bar \3 _f \otimes \3 _f = \8 _f \oplus \1 _f ~\hskip 2cm SU(3)_f,\ee
which are nonets; the flavor sextet representation  of the diquark would have led to flavor multiplets larger than nonets, which as noted above are not observed. 

Therefore, our building-block diquarks  must be those that  are in an antisymmetric configuration under flavor exchange. It is now natural to take this flavor-antisymmetry to be the case not just in the light meson sector but also when we include heavy flavors. For $SU(4)$ flavor, which includes the charm quark, the diquarks form the representations
\be \Q = qq : \hskip .4cm \4 _f \otimes \4 _f = {\bf 10}_f\oplus \bar \6 _f~\hskip 2cm SU(4)_f,\ee
where the ${\bf 10}_f$ is symmetric and the $\bar \6 _f$ is antisymmetric under flavor exchange. The diquark building blocks live in  the antisymmetric $\bar \6 _f$.

\begin{table} 
\renewcommand{\baselinestretch}{1.4}   
\normalsize    

\addcontentsline{lot}{subsection}{Table 1: Diquark configurations}
\renewcommand{\baselinestretch}{1.1}
\small

{\bf 
Table 1: Diquark configurations} (adopted from Jaffe \cite{Jaffe:1999ze}, and adding the $SU(4)_f$ and ${\cal H}_{CE}$ columns). "A" and "S" stand for "Antisymmetric" and "Symmetric" representations, respectively. For a discussion of ${\cal H}_{CM}$ and ${\cal H}_{CE}$, see Section \ref{interquark}.

\renewcommand{\baselinestretch}{1.4}
\normalsize

\[ 
\begin{array}{|c||c c| c| c || c|c|}
\hline  
 & \multicolumn{2}{c|}{Flavor} & Spin & Color&  {\cal H}_{CM} & {\cal H}_{CE} \\
          &SU(3)_f&SU(4)_f                                &SU(2)_s&SU(3)_c&&\\ \hline \hline
\Q_1&\bar \3 _f (A)&\bar \6 _f(A)&\1_s(A) &\bar \3 _c(A)& -8&-8/3\\ \hline
\Q_2& \bar \3 _f(A)&\bar \6_f(A)&\3_s(S)&\6_c(S)&-4/3 & 4/3\\ \hline
\Q_3& \6_f(S)&{\bf 10}_f(S)&\3_s(S)&\bar \3 _c(A) &8/3 &-8/3\\ \hline
\Q_4& \6 _f(S)&{\bf 10}_f(S)&\1_s(A)& \6_c(S)&4 &4/3\\ \hline
\end{array}
\]
\renewcommand{\baselinestretch}{1.5}
\normalsize

\end{table}

Now we list in Table 1 (see also \cite{Jaffe:1999ze}) the flavor, spin, and color states of all the totally antisymmetric configurations of two quarks which are in their lowest orbital. For any given row in the table, the product of all three states must be antisymmetric. There are four such configurations, named $\Q_1$, $\Q_2$, $\Q_3$, and $\Q_4$.  The first two  ($\Q_1$ and $\Q_2$) are the flavor--antisymmetric diquark configurations which are the building blocks for mesons. The $\Q _1$ is also antisymmetric under spin and color; it happens to be the one that played a central role as a proposed constituent of pentaquarks in \cite{Jaffe:2003sg} and has come to be known as the "good" diquark. The other flavor-antisymmetric diquark, $\Q _2$, is symmetric under spin and color, and seems to have been unfairly neglected. We will discuss $\Q_3$ in Section \ref{b-diq} as a building block for baryons; $\Q_4$ does not qualify as a building block for any baryon.

\subsection{Meson quantum numbers}

We now have three types of building blocks for mesons: the quarks $q$ and the two flavor--antisymmetric diquarks $\Q _1$ and $\Q _2$.
Armed with these, we now work out the meson quantum numbers  that we can expect to arise. See Tables 2a, 2b, and 2c.

\begin{table} 
\addcontentsline{lot}{subsection}{Table 2a: $\qq$ meson quantum numbers}
\addcontentsline{lot}{subsection}{Table 2b: $\QQA$ meson quantum numbers}

\renewcommand{\baselinestretch}{1.1} 
\small

{\bf Tables 2a, 2b, 2c: Meson quantum numbers for $\qq$, $\QQA$, and $\QQB$  up to $L=3$}.
The third column  is derived using equations (\ref{jpcqq}) and (\ref{jpcQQ}). 
A $\surd$ indicates that at least one member of the corresponding light nonet has been observed (see Tables 3a, 3b); a dot "$\bullet$" indicates that this nonet consists  mainly of  well-established mesons.  
\vskip .4cm
\renewcommand{\baselinestretch}{1.3} 
\small

\vskip .4cm

\[
\begin{array}{|c|c|c|c||l|}
\hline 
\multicolumn{5}{|c|}{ \mbox{{\bf Table 2a: $q\bar q$  }}}\\
\hline \hline
L&S&J^{PC} &^{2S+1}L_J& \\
\hline \hline
0&0&0^{-+} & ^1S_0&\surd \bullet \\\hline
0&1&1^{--} &^3S_1&\surd \bullet\\ \hline \hline
1&0&1^{+-} &^1P_1&\surd \bullet\\ \hline
1&1&2^{++}&^3P_2&\surd \bullet\\
&&1^{++}&^3P_1&\surd \bullet\\
&&0^{++}&^3P_0&\surd \bullet \\ \hline \hline
2&0&2^{-+}&^1D_2&\surd \bullet\\ \hline
2&1&3^{--}&^3D_3&\surd \bullet\\
&&2^{--}&^3D_2&\surd\\
&&1^{--}&^3D_1&\surd \bullet\\ \hline \hline
3&0&3^{+-}&^1F_3& \surd \\ \hline
3&1&4^{++} & ^3F_4&\surd \bullet\\ 
&&3^{++}&^3F_3&\\
&&2^{++}&^3F_2&\surd \bullet \\ \hline
\multicolumn{5}{c}{ } \\
\multicolumn{5}{c}{ } \\
\multicolumn{5}{c}{ } \\
\hline 
\multicolumn{5}{|c|}{ \mbox{{\bf Table 2b: $\QQA$}}}\\
\hline \hline
L&S&J^{PC} &^{2S+1}L_J&\\ \hline \hline
0&0&0^{++}& ^1S_0&\surd \bullet\\ \hline
1&0&1^{--}&^1P_1&\surd \bullet\\ \hline
2&0&2^{++}&^1D_2&\surd\\ \hline
3&0&3^{--}&^1F_3&\surd\\ 
\hline
\multicolumn{5}{c}{ } \\
\multicolumn{5}{c}{ } \\
\multicolumn{5}{c}{ } \\
\multicolumn{5}{c}{ } \\
\multicolumn{5}{c}{ } \\
\end{array}
\addcontentsline{lot}{subsection}{Table 2c: $\QQB$ meson quantum numbers}
\renewcommand{\baselinestretch}{1.3} 
\small
\hskip 2cm
\begin{array}{|l|l|c|c||l|} 
\hline
\multicolumn{5}{|c|}{ \mbox{{\bf Table 2c: $\QQB$}}}\\
\hline \hline
L&S&J^{PC}&^{2S+1}L_J &\\ \hline \hline
0&0&0^{++}& ^1S_0& \\
0&1&1^{+-} &^3S_1&\\
0&2&2^{++}&^5S_2 &\\ \hline \hline
1&0&1^{--}&^1P_1&\\ \hline
1&1&2^{-+}&^3P_2& \surd \bullet\\
 && 1^{-+}&^3P_1&\surd \\
&&0^{-+}&^3P_0&\surd \bullet\\ \hline
1&2&3^{--} &^5P_3&\\
 & &2^{--}&^5P_2&\surd \bullet \\
&&1^{--}&^5P_1 & \surd \\ \hline \hline
2&0&2^{++}&^1D_2& \surd \\ \hline
2&1&3^{+-}&^3D_3& \\
&& 2^{+-}&^3D_2& \\
&& 1^{+-}&^3D_1&\surd \\ \hline
2&2&4^{++}&^5D_4&\\
&& 3^{++}&^5D_3&\\
&& 2^{++} &^5D_2& \surd\\
&&1^{++}&^5D_1&\surd \\
&&0^{++}&^5D_0&  \surd\\ 
\hline \hline
3&0&3^{--}&^1F_3&\surd\\ \hline
3&1&4^{-+}&^3F_4&\surd  \\
&&3^{-+}&^3F_3 &\\
&&2^{-+}&^3F_2 &\surd \\ \hline 
3&2&5^{--}&^5F_5&\\
&& 4^{--}&^5F_4& \\
&& 3^{--}&^5F_3&\\
&&2^{--} &^5F_2&\\
&&1^{--}&^5F_1& \\ \hline 
\end{array}
\]
\normalsize
\end{table}

\subsubsection*{Color}

Mesons, like all hadrons,  must be color singlets. There are three combinations of our building blocks that yield such objects:  
\bea \qq \; :&&  \3 _c \otimes \bar \3 _c =  \8 _c \oplus \1 _c ~ \hskip 2cm SU(3)_c,\\
\QQA \; :&&\bar \3 _c \otimes \3 _c =\8 _c \oplus \1 _c~ \hskip 2cm SU(3)_c,\\
\QQB \; : && \6 _c \otimes \bar \6 _c = {\bf 27} _c \oplus \8 _c \oplus \1 _c~ \hskip .8cm SU(3)_c,
\eea
 so we have three types of mesons corresponding to the three appearances of $\1 _c$.

\subsubsection*{Flavor}

As we ensured in Section \ref{afew}, in the light quark sector with an $SU(3)$ flavor group all our  mesons -- $\QQA$ and $\QQB$ as well as $\qq$ -- live in flavor nonets: 
\bea \qq \; :&&\3 _f\otimes \bar \3 _f=  \8 _f\oplus \1_f \hskip 2cm SU(3)_f, \\ 
 \QQi \; : && \bar \3 _f\otimes \3 _f=\8 _f\oplus \1_f ~ \hskip 2cm SU(3)_f,
\eea
where $\QQi$ denotes both $\QQA$ and $\QQB$. 

 When we include the charm quark, the flavor group is $SU(4)$, and the $\qq$ lives in 
 \be \qq\; : \hskip 1cm \4 _f\otimes \bar \4 _f= {\bf 15}_f\oplus \1 _f~\hskip 2cm SU(4)_f,\ee 
 while $\QQi$ live in 
 \be \QQi \; : \hskip 1cm \bar \6 _f\otimes \6 _f= {\bf 20}_f\oplus {\bf 15}_f\oplus \1 _f ~ \hskip .8 cm SU(4)_f.\ee 
 
\no These are the flavor multiplets of our mesons. 
 
 \subsubsection*{Spin, parity, and charge}

The total spin $J$, parity $P$, and charge $C$ quantum numbers are denoted $\jpc$. The total spin $J$ is given as usual by adding orbital ($L$) and internal spin ($S$) angular momenta:
\be J=L\otimes S ~. \ee 
The parity  and charge  quantum numbers  are  given by:
\bea \label{jpcqq}\qq \; : &&\hskip 1.2cm  P=(-1)^{L+1}~, \hskip 1cm C=(-1)^{L+S} \\
\label{jpcQQ} \QQi \; : &&\hskip 1.2cm  P=(-1)^L ~,\hskip 1.3cm C=(-1)^{L+S}~.
\eea

One should note that the orbital angular momentum $L$ is  between the quark and antiquark or diquark and antidiquark; the internal orbital angular momentum of our building blocks is zero.

We list all the allowed $\jpc$ quantum numbers along with their corresponding $L$ and $S$ for $L\leq 3$ in Tables 2a ($\qq$), 2b ($\QQA$), and 2c ($\QQB$).

\section{Re--classification of Mesons}\label{reclass}

We are now ready to re-classify the meson spectrum. We carry out the following procedure{\bf :} we compile a list -- from the Particle Listings in the PDG \cite{Beringer:1900zz} --  of all the mesons\footnote{We omit those listed under "further states" in the PDG, as they have not been confirmed.} along with their  masses, flavor and $\jpc$ quantum numbers. We arrange them into flavor multiplets with approximately degenerate masses and common $\jpc$; the light mesons form either full or partial flavor nonets.
Then, we turn to our tables of meson quantum numbers (Tables 2a, 2b, 2c) for all the occurrences of each $\jpc$
 and assign each multiplet (or partial multiplet) to an appropriate 
meson type ($\qq$, $\QQA$, or $\QQB$) with specified $L$ and $S$ quantum numbers.

In making the assignment, we make a rough but standard assumption \cite{Jaffe:1999ze,Rosenzweig:1975fq,Burns:2004wy,Isgur:1999jv}, which we call the {\it "orbital excitation rule,"} that every unit of orbital angular momentum $L$ contributes about $.5$GeV to the mass of a light meson.
  Therefore,  roughly speaking  we expect the following mass ranges for light mesons:
\bea \non &\mbox{S--waves} \hskip 1cm & m \leq 1 GeV \\ 
\label{Lexc} \mbox{{\it orbital excitation rule:}} \hskip 1.5cm &\mbox{P--waves} \hskip 1cm & 1 < m < 1.5  GeV \\ 
\non &\mbox{D--waves} \hskip 1cm  & 1.5 < m <  2 GeV \\ 
\non & \mbox{F--waves}\hskip 1cm & 2 < m < 2.5 GeV ~, 
\eea 
 and so on.

This procedure produces the following tables:  Table 3a (light mesons), Table 3b (charmed and bottom mesons), and Table 3c (isorons and magic $\jpc$, both light and heavy, to be defined below).\footnote{For a multiplet by multiplet discussion of the process, see Appendix \ref{details}.}

We now analyze the outcome.

\begin{table}

\renewcommand{\baselinestretch}{1.3}   
\normalsize    

\addcontentsline{lot}{subsection}{Table 3a: Assignments for light mesons}

\renewcommand{\baselinestretch}{1}   
\footnotesize

{\bf Table 3a: Our suggested assignments for observed light mesons}. Compare with Table 14.2 in the PDG \cite{Beringer:1900zz}. The second and third columns are the quark/diquark constituents and their orbital and spin quantum numbers, respectively. 
A blank space in the fourth column indicates that the anticipated meson has not yet been detected. 
A dot "$\bullet$"  next to a meson indicates it is considered well--established by the PDG. See Appendix A for a line-by-line discussion. 
 
 \renewcommand{\baselinestretch}{1.2}   
\normalsize

 \footnotesize

\[
\begin{array}{|c|c|c|lll|}
\hline
J^{PC}& \mbox{constituents}& ^{2S+1}L_J& I=1& I={1\over 2} & \hskip 1cm I=0 \\
\hline \hline
0^{-+}& \qq & ^1S_0& \bullet \pi  & \bullet K &\bullet \, \eta  \hskip 1.3cm \bullet \eta '(958) \\
\hline
0^{-+}& \QQB & ^3P_0&\bullet \pi (1300) & K(1460) & \bullet \, \eta (1475) \; \; \; \bullet \eta (1295) \\
\hline
\hline 
0^{++} &\QQA & ^1S_0& \bullet a_0(980) & \kappa (800) & \bullet f_0(980) \; \;  \; \; \bullet f_0(600)  \\
\hline
0^{++}& \qq& ^3P_0 & \bullet a_0 (1450) & \bullet K_0^*(1430) & \bullet f_0(1710)\; \; \bullet f_0(1370) \\ \hline
0^{++}&\QQB &^5D_0&&K_0^*(1950)& f_0(2100)\; \; \; \; \bullet f_0(2020)   \\
\hline \hline
1^{--}& \qq & ^3S_1& \bullet \rho (770) & \bullet K^*(892) &\bullet \,  \phi (1020)\;  \; \; \bullet \omega (782)  \\
\hline
1^{--}&\QQA& ^1P_1& \bullet \rho (1450)& \bullet K^*(1410) &  \bullet \,  \phi (1680) \; \; \;  \bullet  \omega (1420)\\
\hline
1^{--}&\QQB& ^5P_1&\rho(1570)&& \\ \hline
1^{--}& \qq & ^3D_1 & \bullet \rho(1700)& \bullet K^*(1680) & \hskip 1.8cm \bullet \, \omega(1650)  \\ \hline
1^{--}&\QQB&^5F_1&\rho(2150)&& \phi(2170)\\ \hline
\hline 
1^{-+}& \QQB & ^3P_1 & \bullet \pi _1(1600) & K(1630) &   \\ 
\hline \hline
1^{++}&\qq  &
^3P_1 &\bullet a_1(1260)& \bullet K_1(1400) &  \bullet f_1(1420) \; \bullet f_1(1285) \\ \hline
1^{++}& \QQB & ^5D_1& a_1 (1640) & K_1(1650)&  f_1(1510)  \; \\
\hline \hline
1^{+-}& \qq & ^1P_1& \bullet b_1(1235)& \bullet K_1(1270)  &  h_1(1380) \; \; \bullet h_1(1170) \\ \hline
1^{+-}&\QQB&^3D_1&  &  & h_1(1595) \\ \hline
\hline
2^{-+} &\QQB &  ^3P_2& \bullet \pi _2 (1670)&  K_2(1580)  & \eta _2 (1870) \; \; \;  \bullet \eta _2 (1645) \\ \hline
2^{-+}&   \qq    &  ^1D_2  & \bullet \pi_2(1880) &                                         & \\ \hline
2^{-+}& \QQB &^3F_2  & \pi _2(2100) & K_2(2250) & \\
\hline \hline
2^{--}&\QQB &^5P_2 && \bullet K_2(1770)&  \\ \hline
2^{--}& \qq&  ^3D_2& & \bullet K_2(1820)&   \\
\hline 
\hline
2^{++}& \qq & ^3P_2&\bullet a_2(1320) &\bullet K_2^*(1430) &  f_2(1430)  \; \; \; \;  \bullet f_2(1270) \\ \hline
2^{++}& \QQB& ^1D_2& &&\bullet f_2'(1525) \\ \hline 
2^{++}& \QQA &^1D_2 & a_2(1700) && f_2(1640) \; \; \; \;  \; \; f_2(1565) \\ \hline 
2^{++}&\QQB& ^5D_2&&&f_2(1810) \\ \hline 
2^{++}&  \qq & ^3F_2& & K_2^*(1980) & \bullet f_2(2010)  \; \;  \;  \bullet f_2(1950) \\ 
\hline \hline 
3^{--}&\qq &^3D_3& \bullet \rho _3(1690)& \bullet K_3(1780) & \bullet \, \phi _3 (1850) \; \bullet  \omega _3(1670) \\ \hline
3^{--}&\QQA&^1F_3&\rho_3 (1990)&& \\ \hline
3^{--}&\QQB&^1F_3&\rho_3 (2250)&&  \\ \hline
 \hline 
 3^{+-}&\qq&^1F_3& &K_3(2320) &  \\ \hline \hline 
 4^{-+}&\QQB&^3F_4& &K_4(2500)& \\ \hline \hline 
4^{++}&\qq &  ^3F_4& \bullet a_4(2040) & \bullet K_4^*(2045) &  f_4(2220) \;\; \;  \bullet f_4(2050) \\ \hline
\hline
5^{--}&\QQB &^5F_5 & \rho _5 (2350) & K_5^*(2380)&  \\
\hline \hline
6^{++}&\QQB & ^5G_6 &a_6(2450) && f_6(2510) \\
\hline 
\end{array}
\]
\end{table}

\begin{sidewaystable}


\renewcommand{\baselinestretch}{1}   
\small

{\bf Table 3b: Our suggested assignments for observed heavy (charm and bottom)  mesons.}
Compare with Table 14.3 in the PDG \cite{Beringer:1900zz}. We include the new 
$Z_b$ states.
 We use the mesons' names as they appear in the PDG for convenience, without agreeing with the radial or orbital assignments that sometimes appear in a meson's name.
The second and third columns are the quark/diquark constituents and their orbital and spin quantum numbers, respectively.
A blank space in the fourth column indicates that the anticipated meson has not yet been detected. 
A dot "$\bullet$"  next to a meson indicates it is considered well--established by the PDG. See Appendix A for a line-by-line discussion. 

\renewcommand{\baselinestretch}{1.3}   
\normalsize

\vskip .2cm
\addcontentsline{lot}{subsection}{Table 3b: Assignments for heavy (charm and bottom) mesons}
\small
\[ 
\begin{array}{|l|c|c|llll|llll|}
\hline 
&&&\multicolumn{4}{|c|}{\mbox{{\bf Charmed mesons}}}&\multicolumn{4}{|c||}{\mbox{{\bf Bottom mesons}}}\\
\jpc & \mbox{constituents}& ^{2S+1}L_J&I=1^{\circ}&I={1\over 2} &I=0 &I=0 &I=1^{\circ}&I={1\over 2} &I=0  &I=0 \\
\hline \hline
 0^{-+} &\qq&^1S_0&&\bullet D  & \bullet D_s ^{\sharp}& \bullet \eta _c (1S) &&\bullet B ^{\dagger }&\bullet B_s ^{\dagger },  B_c^{\dagger }&\eta _b (1S)^{\dagger} \\ \hline
0^{-+}&\QQB &^3P_0&&&& \bullet \eta _c (2S)^{\dagger} &&&& \\
\hline \hline
 0^{++} &\QQA &^1S_0&&D_0^*(2400)^{\sharp}  & \bullet D_{s0}^*(2317)& \bullet \chi _{c0}(1P) &&&&\bullet \chi _{b0}(1P) \\ \hline 
0^{++}&\qq &^3P_0&&&&&&&&\chi _{b0}(2P)^{\dagger \dagger} \\
\hline \hline
 1^{--} &\qq&^3S_1  &&\bullet D^* & \bullet D_s^{* \sharp} & \bullet J/\psi (1S)&&\bullet B^{*\dagger }&B_s^{*\dagger}&\bullet \Upsilon (1S) \\ \hline 
 1^{--} &\QQA&^1P_1  & && & \bullet \psi (2S)&&&&\bullet \Upsilon (2S) \\ \hline 
 1^{--} &\qq& ^3D_1 && & &  \bullet \psi(3770)&&&&\bullet \Upsilon (3S) \\ \hline 
 1^{--} &\QQB&^5F_1&&  & & \bullet \psi (4040)&&&&\bullet \Upsilon (4S) \\ \hline \hline
 1^{++} &\qq&^3P_1&& D_1(2420)& \bullet D_{s1}(2536)^{\sharp}& \bullet \chi _{c1}(1P) &&\bullet B_1(5721)^{0\dagger}& \bullet B_{s1}(5830)^{0\dagger}&\bullet \chi _{b1}(1P)^{\dagger \dagger}\\  \hline 
1^{++}&\QQB&^5D_1 && &\bullet D_{s1}(2460)& \bullet X(3872)^{\sharp \sharp}& Z_b(10610)^{\sharp \sharp \sharp } &&& \bullet \chi_{b1}(2P)^{\dagger \dagger}\\
\hline \hline
2^{++} &\qq&^3P_2 &&\bullet D_2^*(2460)&  \bullet D_{s2}(2573)^{\sharp}& \bullet \chi _{c2}(1P)&& \bullet B_2^*(5747)^{0\dagger} & \bullet B_{s2}^*(5840)^\dagger &\bullet \chi_{b2}(1P)^{\dagger \dagger} \\ \hline 
2^{++}& \QQA &^1D_2&&&& \chi_{c2}(2P)&&&&\bullet \chi_{b2}(2P)^{\dagger \dagger} \\ 
\hline 
\end{array}
\]

$^{\circ}$ $I=1$ applies only to $\QQi$ multiplets; no $I=1$ is expected in charm or bottom $\qq$ mesons. 

$^{\dagger}$ $I$, $\jpc$ need confirmation

$^{\dagger \dagger}$ J  needs confirmation. 

$^{\sharp}$ $J^P$ needs confirmation. 

$^{\sharp \sharp}$ Quantum numbers not established; the $X(3872)$ mixes isospin 0 and 1 (see Appendix A.2).

$^{\sharp \sharp \sharp }$ The state $Z_b(10610)$, and the related state $Z_b(10650)$, had not yet been detected by BELLE \cite{Belle:2011aa}  
 when the original version of this paper was posted on arXiv \cite{arxiv}. 

\end{sidewaystable}

\normalsize

\begin{table}

\renewcommand{\baselinestretch}{1}   
\small

\addcontentsline{lot}{subsection}{Table 3c: Isorons and magic $\jpc$}
{\bf Table 3c: Isorons and magic $\jpc$}  (see Section \ref{isoronsmagic}). The magic $\jpc$ are $0^{-+}$, $0^{++}$, $2^{++}$ for light isorons and $1^{--}$ for heavy isorons. A dot "$\bullet$"  next to a meson indicates it is considered well--established by the PDG. In the glueball table at the bottom, a column with an "X" indicates that the corresponding $\jpc$ quantum number is expected for glueballs by lattice QCD. 

\renewcommand{\baselinestretch}{1.3}   
\small

\[
\begin{array}{|c|c|c|c|c|c|c|c|}
\hline
\multicolumn{8}{|c|}{\mbox{\small{{\bf Isorons}}}} \\ 
\hline
\hskip 1cm&0^{-+}   &0^{++}   &1^{--}     &1^{-+}  &1^{++}   &2^{++}   &4^{++}       \\ \hline
 \multicolumn{7}{|c}{  } &\\ \hline
&\bullet \eta(1405)&\bullet f_0(1500)&\rho(1900) &\bullet \pi _1 (1400) & & f_2(1910)&f_4(2300)\\
&\eta(1760)    & f_0(2200)  &  & &    &f_2(2150)    & \\
\mbox{\small{Light}}&\bullet \pi (1800)&f_0(2330) &&  & &\bullet f_2(2300)& \\
&K(1830)&&&&&\bullet f_2(2340)&\\
&\eta(2225) & &&&&&\\ \hline
 \multicolumn{7}{|c}{  } &\\ \hline
& &&\bullet \psi (4160)& &Z_b(10650)^{\sharp \sharp \sharp }&& \\
&&& \bullet X(4260)&&&& \\
\mbox{\small{Heavy}}&&& X(4360)&&&& \\
&&& \bullet \psi (4415)&&&&\\
&&&\Upsilon (10860)&&&&\\
&&&\Upsilon (11020)&&&&\\ 
\hline 
 \multicolumn{7}{c}{  } \\ \hline
\multicolumn{8}{|c|}{\mbox{\small{{\bf Glueballs}}}} \\ 
\hline
&0^{-+}   &0^{++}   &1^{--}     &1^{-+} &1^{++}   &2^{++}   &4^{++}       \\ \hline
&X&X&&&&X& \\ \hline
\end{array}
\]
\vskip .4cm
$^{\sharp \sharp \sharp }$See footnote in Table 3b
\end{table}

\normalsize

\subsection{Isorons, magic numbers, and glueballs}\label{isoronsmagic}

In most cases, our procedure above resulted in a unique assignment for each meson, which appears in Tables 3a and 3b. 
As we were carrying out the procedure, we noticed that sometimes, multiple mesons which carry the same quantum numbers but different masses vied for one available space in the tables. One of these, usually the one most closely degenerate in mass with the relevant multiplet, was placed in that available space. The others are hereby named {\it isorons}, short for iso-hadrons and analogous to isotopes of atomic physics. Recall the standard definition for isotopes\footnote{Definition taken from Encyclopaedia Brittanica online.}:

\renewcommand{\baselinestretch}{1.1}  
\normalsize

\begin{quote}

"any of two or more species of atoms of a chemical element with the same atomic number and position in the periodic table and nearly identical chemical behavior but with differing atomic mass or mass number and different physical properties." 

\end{quote}
\renewcommand{\baselinestretch}{1.4}  
\normalsize

\no Just as with isotopes, we define an isoron to be one of two or more species of mesons with the same quantum numbers  but different mass. 
The isorons are an integral part of the hadronic spectrum, the same way that isotopes are an integral component of the elements.
The isorons are listed in Table 3c  by $\jpc$.

There are certain $\jpc$'s for which there is an abundance of isorons; we name these "magic $\jpc$" in analogy with the magic numbers of the nuclear shell model \cite{shell}. From Table 3c we see that the magic $\jpc$ are $0^{-+}, \; 0^{++}, \; 2^{++}$  for light mesons and $1^{--}$ for heavy mesons. 

 Strikingly, the magic $\jpc$ for light mesons match the $\jpc$ expected for low-lying glueballs: lattice QCD calculations indicate that ground state glueballs have $\jpc =0^{++}$ and the first two excited states of glueballs have $\jpc =2^{++}$ and $\jpc = 0^{-+}$ \cite{Morningstar:1999rf,Chen:2005mg}. This matching cannot be a coincidence -- there must be a deep underlying reason for it. That reason is beyond the scope of this paper.
 
\subsection{No radials}\label{norad}

As we can see, a central result of our classification is that there is no radial quantum number, which indicates that there are no radially excited mesons. 
 The meson multiplets which have been believed\footnote{We take Tables 14.2 and 14.3 of the PDG \cite{Beringer:1900zz} to be the currently accepted quark model classification.} to be radially excited $\qq$  
 are orbitally excited $\QQi$:
\begin{itemize}
\item the second $0^{-+}$ nonet, which was classified in the literature and in Table 14.2 of the PDG as a radial excitation with $\nslj = 2^1S_0$,  finds its place here as a $\QQB$ with $\slj=\; ^3P_0$;
\item the second $1^{--}$ nonet, which was classified in the literature and in Table 14.2 of the PDG as a radial excitation with  $\nslj = 2^3S_1$, finds its place here as a $\QQA$ with $\slj=\; ^1P_1$.
\end{itemize}

As we show later on (in Section \ref{baryons}), there is no radial quantum number in the baryon sector either, so put together, there are no radial excitations in the entire hadron spectrum. 

But, how can we reconcile our classification and its results with the fact 
that for so many years it has been believed that radial excitations of hadrons do exist? 

One of the main sources for the concept that hadrons may be radially excited  
goes back to potential models. According to these models,
 low--energy QCD is described by a quark--quark potential $V(r)$, where $r$ is the distance between the quarks. The potential in these models has two terms: a short--distance term that is Coulomb-like (i.e., proportional to $-1/r$) and analogous to the interaction between the proton and electron in the hydrogen atom, and a long--distance term $V_{conf}(r)$ that increases with $r$ and -- according to the models -- describes confinement. (For a review of potential models, see \cite{Lucha:1991vn}.)


In these models, the spectrum for quark--antiquark bound states, i.e. mesons, is obtained by solving the Schr\"{o}dinger equation with the above potential $V(r)$.
As with the hydrogen atom, or  as with any central potential in non-relativistic quantum mechanics, the resulting quantum numbers that describe the spectrum include
a principal or radial quantum number $n$. Hence, potential models automatically allow for, and in fact require, radial quantum numbers and radial excitations. Other studies of QCD have also employed analogies with the hydrogen atom; for a recent example see \cite{Brodsky:2008be}.

In contrast, 
 in the early versions of the PDG \cite{OLDRPP}, 
starting in the 1960's when the quark model was first proposed, 
mesons were classified only by spin and orbital quantum numbers: 
\be \slj \; . \ee 
There was no radial quantum number $n$.
Similarly, early discussions of the quark model did not mention radial excitations or a radial quantum number \cite{Jaffe:1977xy}. The quark model certainly does not call for a radial quantum number.  Radial quantum numbers for the hadron spectrum appeared in the PDG for the first time only  in 1980 \cite{Kelly:1980kr}. The atomic notation 
\be n\slj ~,\ee 
which includes the radial quantum number $n$, was adopted by the PDG for the hadron spectrum only in 1992 \cite{Hikasa:1992je}.
Interestingly enough,  the classification of some mesons as radials in the  PDG's from 1992 through 2002 was partially retracted in the subsequent versions (compare Table 13.2 of \cite{Hagiwara:2002fs} to Table 14.2 of \cite{PDG04} or \cite{Beringer:1900zz}): their classification as radials was considered far-fetched \cite{privateI,privateII}.

Was there ever any {\it experimental} evidence for  a radial quantum number in hadrons? 
As of now, the internal radial structure of hadrons has not  been experimentally probed: 
all that has been reported so far  is a measurement of the form factors of a few low--lying hadrons, from which their charge radius can be inferred (this has been done for $\pi ^\pm$, $K^\pm$, $p$, $\Sigma ^-$) \cite{Beringer:1900zz}. So the radial quantum number that ultimately crept into the quark model classification tables and the PDG was actually 
an artifact of the models rather than 
a quantity arising from any measured property of hadrons or quarks.

Furthermore, theoretical predictions about radial excitations in hadrons have been known to encounter difficulties: data involving the masses of the candidates for radial excitations shows that they are often significantly lighter  than predicted by the models,
 and data involving their decay modes often does not favor a radial assignment either \cite{RadialModels}. 

In retrospect, it is actually natural that the quantum numbers of hadrons are different from those of atoms. After all, the hydrogen atom, and the entire atomic system, is inherently different from low--energy QCD even if only because atoms are ionizable whereas low--energy QCD is confining.

We leave a more complete discussion of the implications of the result that there are no radially excited hadrons to our companion paper \cite{companionII}.

\subsection{No "exotics" or other outcasts}\label{noexot}
The traditional quark model allows only for $\qq$ mesons. The term "exotic meson" refers to those mesons which do not fit into the traditional quark model. While for many years  there seemed to be a very small number of exotic mesons -- the light scalar mesons were the only ones unexpected by the model -- more and more exotic mesons have recently been discovered, including several charmed mesons and a few pions. None of these mesons can be adequately explained within the traditional quark model. 

Our model embraces these mesons as legitimate constituents of the hadron spectrum, and they are no longer "exotic." Instead, they are made up of $\QQi$.  
These formerly exotic mesons, along with their classification, include:
\begin{itemize}
\item the "cryptoexotic" \cite{Jaffe:2004ph} light scalar nonet with $\jpc =0^{++}$ is  a $\QQA$, $ ^1S_0$ (see also \cite{Jaffe:1976ig});
\item a manifestly "exotic" meson with $\jpc =1^{-+}$ is  a $\QQB$, $ ^3P_1$ (see also \cite{Chung:2002fz});
\item some newly discovered charmed mesons, (see \cite{CHARM}) 
including:
\begin{itemize}
\item the $D_{sJ}^*(2317)$ with $\jpc = 0^{++}$ is a $\QQA$, $^1S_0$; 
\item the $D_{sJ}(2460)$ with $\jpc = 1^{++}$ is a $\QQB$, $^5D_1$;
\item the $X(3872)$ with $\jpc =1^{++}$ is a $\QQB$,$^5D_1$.
\end{itemize}
\end{itemize}

\vskip .2cm
\no There are also numerous other mesons which have been just left out of the classification tables of the traditional quark model -- see Appendix B, Table 5 for a complete list of the unclassified mesons.  These are also embraced into our model, for example:
\begin{itemize}
\item some heavier scalar mesons with $\jpc =0^{++}$ now form a nonet which is classified as  $\QQB$, $ ^5D_0$;
\item some vector mesons with $\jpc = 1^{++}$ are  now $\QQB$, $ ^5D_1$;
\item some $2^{++}$ mesons which are now $\QQA$, $ ^1D_2$.
\end{itemize}

\subsection{New particles}\label{newpart}

Our model implies the existence of new particles. Any blank space in Tables 3a and 3b represents a meson that we anticipate will be detected. In addition, any row in Tables 2a, 2b, and 2c 
which does not have a $"\surd"$ in the rightmost column represents an anticipated multiplet. 

In the PDG \cite{Beringer:1900zz} there is a list of light "further states,"  which are "states observed by a single group or states poorly established." We did not use these mesons in our study, but
quite a few of the blank spaces in the tables may be filled  by these mesons if they are eventually confirmed. For example, the $\omega (1960)$ may partially complete the fifth $1^{--}$ nonet; the $\rho _2 (1940)$ and the $\omega _2(1975)$ may partially complete the second $2^{--}$ nonet; the $a_2(1990)$ may complete the third $2^{++}$ nonet; the $\omega _3(1945)$ and $\omega _3 (2255)$ may partially complete the second and third $3^{--}$ nonet, respectively; the $b_3(2030)$, $h_3(2025)$, and $h_3(2275)$ may complete the first $3^{+-}$ nonet; and $\omega _5(2250)$ may partially complete the $5^{--}$ nonet. 

Other mesons whose detection would support our model are those whose quantum numbers are part of our model but are prohibited in the traditional quark model (these are the "manifestly exotic" quantum numbers). These include $\jpc=1^{-+}, 2^{+-}, 3^{-+}$, etc.; these are manifestly exotic with respect to the quark model but they appear in our model as $\QQB$. Some $\jpc=1^{-+}$ mesons (the $\pi _1(1400)$ and $\pi _1(1600)$) have already been established and their existence is  evidence already supporting our model. 
It is interesting that the $\jpc = 1^{-+}$ pions were detected relatively recently (in 1997 \cite{Thompson:1997bs}), and in fact acquired well--established status in the PDG only in 2004; we believe their cousins with $\jpc = 2^{+-},3^{-+}$ will follow suit.

\subsection{Mass hierarchies in light nonets}\label{hierarchy}

A strange quark is heavier than an up or down quark. Therefore, in the  light meson sector, a strange quark constituent makes a meson heavier.  As a result,  the mass hierarchy within a $\QQi$  nonet is expected to be inverted as compared to the mass hierarchy of a  $\qq$ nonet \cite{Jaffe:1976ig,Jaffe:1999ze,Black:1998wt}. 
That is,  in $\qq$ nonets, the $I=1/2$ mesons (one strange quark) are heavier than the $I=1$ mesons (no strange quarks), while in $\QQi$ nonets the $I=1/2$ mesons (one strange quark) are lighter than the $I=1$ mesons (two strange quarks). 
This is particularly prominent for the first $0^{++}$ nonet, whose mass hierarchy is clearly inverted, as was first noted in \cite{Jaffe:1976ig}. 

The results obtained through our classification are consistent with this expected mass hierarchy in almost all cases. However, it should be noted that sometimes, the actual mass hierarchies cannot be read off from Table 3a. For one, the names of the mesons do not always reflect the meson's mass:  generally, a meson's name in the PDG does not get updated when mass measurements are improved, sometimes making it appear as though the mass hierarchy in a nonet is the opposite of what it really is. For example, in the $1^{-+}$ nonet, classified as a $\QQB$, the mass of the $\pi _1(1600)$ is actually 1662 MeV\footnote{This mass was reported as 1596 MeV in earlier editions of the PDG.}, 
 and the mass of the $K(1630)$ is 1629 MeV, so the $K(1630)$ is in fact lighter than the $\pi_1(1600)$, making the mass hierarchy inverted as expected. 

Also, there are experimental errors in mass measurements that are significant and could make the mass hierarchy of a nonet uncertain. In our classification, in the second $1^{++}$ nonet, the mass of the $a_1(1640)$ is actually $1647\pm 22$ MeV and the mass of the $K_1(1650)$ is $1650\pm 50$ MeV, so it could very well be that the $K_1(1650)$ is lighter than the $a_1(1640)$, consistent with an inverted hierarchy of a $\QQB$ nonet. 
In the fourth $1^{--}$ nonet, where the $\rho(1700)$ has mass $1720\pm 20$ MeV and the $K^*(1680)$ has mass $1717\pm 27$ MeV, so it could very well be that the $K^*(1680)$ is heavier than the $\rho(1700)$,  consistent with a $\qq$ nonet. 
In the second $0^{-+}$ nonet, the $\pi (1300)$ appears lighter than the $K(1460)$, but the mass of the $\pi (1300)$ is $1300\pm 100$ MeV, and the $K(1460)$ seems to have  been measured only twice over 25 years ago, once giving the mass $1400$ MeV and once giving the mass $1460$ MeV. Therefore, it is possible  that the $K(1460)$ would eventually be found to be lighter than the $\pi (1300)$, consistent with our $\QQB$ assignment. Another example of this kind is the second $0^{++}$ nonet. 

The third $2^{-+}$ nonet is the only one that at this time appears to have an unexpected mass hierarchy.

\subsection{Binding energies of the diquarks}\label{bind}

While many of the  expected $\QQA$ mesons have been observed, the same is not true of the $\QQB$ mesons. This fact alone leads us to believe that the $\Q_2$ is less tightly  bound than the  $\Q _1$. 

We can use our classification to compare the binding energies of the $\Q_1$ and $\Q_2$ diquarks because in 
 the light $\jpc=3^{--}$ sector, we have both a $\QQA$ and a $\QQB$ with the same orbital angular momentum (both are F-waves) and the same isospin. The difference in their masses, which is around $250$MeV, is a rough indication of the difference in binding energies of the $\Q_1$ and $\Q_2$ constituents. Therefore, the binding energy of the $\Q_2$ is roughly
 \footnote{This rough estimate does not take into account the difference between  binding of $\Q _1$ to $\bar \Q _1$ and the binding of $\Q _2 $ to $\bar \Q _2$.}
 $125$MeV less than the binding energy of the $\Q _1$.\footnote{This is consistent with the difference in their binding energies under the interaction ${\cal H}_{CM}$ ($\Delta E = (-8+4/3)\cdot 20$MeV$ = 133$MeV); see  Table 1 and Section \ref{interquark}.}

This implies that the $\Q _2$ is lighter than  the ``bad'' diquark $\Q _3$, which is believed to be about $200-300$MeV heavier than $\Q_1$ \cite{Rosenzweig:1975fq,Wilczek:2004im}.

\subsection{Decays of diquark--antidiquark mesons and the $N\bar N$ threshold}\label{NN}

Our schematic model is not intended to provide detailed predictions about decays of the three types of  mesons in our model.\footnote{As pointed out in \cite{Barnes:1999gv}, the data for decay amplitudes and branching fractions for mesons is anyway far from accurate, making it difficult to test any strong decay models \cite{3P0,Kokoski:1985is,Ackleh:1996yt}.}
However, we can still use our model to say something about these decays.

\label{moredecaydiss} There is a clear distinction between the expected decays of $\QQA$ mesons and $\QQB$ mesons \cite{Chan:1977ty, Jaffe:1977cv, Hendry:1978hv}. This distinction is due to the fact that $\Q _1$ is a color antitriplet ($\bar \3 _c$), while $\Q_2$  is a color sextet ($\6 _c$). 

When a quark--antiquark pair is produced from the vacuum, the quark -- which is a color triplet ($\3 _c$) -- can join the diquark $\Q _1$ to form a baryon, since their tensor product contains a color singlet: 
\be q\Q_1: \hskip 1cm \3 _c\otimes \bar \3 _c = \8 _c \oplus \1 _c  \hskip 2cm SU(3)_c.\ee
Similarly, the antiquark can join the antidiquark to form an antibaryon. When these processes are put together, the quark--antiquark pair joins the $\QQA$ to form a loosely bound baryon--antibaryon molecule which would  dissociate quickly. 

The $\QQB$ is protected from such a  process  since a color sextet cannot join a quark or antiquark to form the color singlet necessary for the formation of a baryon. This can be seen from the absence of $\1 _c$ in the following decompositions:
\bea q\Q_2:\hskip 1cm \3 _c \otimes \6 _c &=& {\bf 10} _c \oplus \8 _c ~\hskip 2cm SU(3)_c,\\
\bar q\Q_2:\hskip 1cm \bar \3 _c \otimes \6 _c &=& {\bf 15} _c\oplus \3 _c ~\hskip 2cm SU(3)_c.
\eea  

Therefore, we would not expect to see $\QQA$ mesons above the nucleon--antinucleon threshold (around 2GeV for light mesons); if any such states do exist, they should be very broad and difficult to detect. On the other hand, $\QQB$ mesons above 2GeV may be narrow. 

Our classification shows (Table 3a)  that indeed, there are no light $\QQA$ above the nucleon--antinucleon threshold. 

\vskip .2cm

 \section{The Baryon Sector}\label{baryons}
 We have stated that there are no radial excitations in the meson spectrum. Can we make an analogous statement about the baryon spectrum? As we show in this section, the answer is "yes."  
Note that we will not  carry out a reclassification of the entire baryon spectrum since in essence this has already been done \cite{Wilczek:2004im,Selem:2006nd,Ida:1966ab,Lichtenberg:1967}.

\subsection{Diquark building blocks for baryons}\label{b-diq}

A baryon, like any hadron, must be a singlet under the color group. If we assume that a baryon is made of a quark and a diquark, then in order for a quark--diquark state to contain a color singlet corresponding to a baryon,
the diquark has to be a color antitriplet $\bar \3 _c$: 
\be  q\Q:\hskip 1cm  \3 _c \otimes \bar \3 _c = \8 _c \oplus \1 _c ~\hskip 2cm SU(3)_c \, .\ee
If the diquark were a color sextet $\6 _c$, combining it with a quark would not result in a  color singlet so no hadron could form:
\be \label{colorbaryon} 
  q\Q: \hskip 1cm  \3 _c \otimes \6 _c = {\bf 10}_c \oplus \8 _c ~\hskip 1.7cm SU(3)_c \, .
\ee

Therefore, the diquark building blocks for the baryon sector are the color--antisymmetric ones, $\Q _1$ and $\Q _3$ (see Table 1 and \cite{Wilczek:2004im,Selem:2006nd,Ida:1966ab,Lichtenberg:1967}). 

The flavor multiplets that can be obtained from $\Q _1$ are octets and singlets  while from $\Q _3$ we obtain decuplets and octets:
\bea q\Q_1 \; : && \hskip 1cm \3 _f \otimes \bar \3 _f = \8 _f \oplus \1 _f ~ \hskip 2.2cm SU(3)_f, \\
q\Q_3\; : && \hskip 1cm \3 _f \otimes \6 _f = {\bf 10}_f \oplus \8 _f ~\hskip 2cm SU(3)_f. \eea

\subsection{Baryons and radials}\label{roper}

 There are only a few baryons that have been believed to be candidates for radial excitations, as classified in the literature and the PDG (Table 14.6 of \cite{Beringer:1900zz}). The lightest one, $N(1440)$, is known as the Roper resonance \cite{DeGrand:1977fy, Rosina:2004sz}. The full list of light ones is:
\bea &&1/2^+: ~ N(1440),~\Lambda (1600),~ \Sigma (1660);  \hskip .3cm \mbox{("Roper octet")}\\
&&1/2^+: ~N(1710),~\Lambda (1810),~\Sigma(1880); \nonumber \\
&&3/2^+: ~\Delta (1600). \nonumber 
\eea

It was  shown in a different context in \cite{Jaffe:2003sg, Selem:2006nd} that  these baryons can be identified with orbitally excited states of the form $\Q _1 \Q _1 \bar q$, where the two $\Q_1$ diquarks are in a relative P--wave. Specifically, the  $N(1440)$, $\Lambda (1600)$, $\Sigma (1660)$, $N(1710)$, and a $\Sigma$ around $1850$MeV  are $\Q _1 \Q _1 \bar q$ states in an $SU(3)_f$ $\8_f \oplus \bar \ten _f$ with
 $J^P=1/2^+$ and $L=1$  ($L$ denotes the relative orbital angular momentum between the $\Q_1$'s) and no radial quantum number. Similarly, we suggest that the $\Delta (1600)$ is a $\Q _1 \Q _1 \bar q$ state belonging to an $SU(3)_f$ $\8_f \oplus \bar \ten _f$ with $J^P=3/2^+$ and $L=1$, again with no radial quantum number.

So, just as in the meson spectrum, there are no radial excitations in the baryon spectrum either.

\section{Interquark Forces in Mesons and Baryons}\label{interquark}

As we noted before, the interquark forces of low--energy QCD are not known, so we ended up deriving the diquark building blocks from 
  meson and baryon phenomenology and properties of color and flavor representations.  
We found that  the diquark building blocks of mesons are flavor-antisymmetric ($\Q _1$ and $\Q _2$ in Table 1), while  the diquark building blocks of baryons are color-antisymmetric ($\Q _1$ and $\Q _3$ in Table 1). 
Since the diquarks in the meson sector are different from the diquarks in the baryon sector, the interquark forces in the meson and baryon sectors must also be different.

It is now natural to seek to learn something about the interquark interactions from these phenomena. 

 As it happens,  there is an interaction under which the attractive diquark configurations are the flavor--antisymmetric ones, $\Q_1$ and $\Q_2$, as in the meson spectrum. That interaction is the spin--dependent part of one gluon exchange (OGE), also known as the color--magnetic interaction ${\cal H}_{CM}$. It was introduced as an important ingredient of hadron spectroscopy in \cite{DeRujula:1975ge}, and  it is given by 
\be \label{CM} {\cal H}_{CM} \propto - \lambda _1 \cdot \lambda _2 \, \sigma _1 \cdot \sigma _2 , \ee
where $\sigma _i$ and $\lambda _i$ are respectively the spin and color operators of the $i$th quark   (the spin--orbit interaction terms of OGE vanish for ground--state diquarks \cite{DeRujula:1975ge}).

The values of ${\cal H}_{CM}$ for each diquark configuration can be obtained by  defining flavor, spin, and color exchange operators $P_f$, $P_s$, and $P_c$, which equal $+1$ if the quarks are symmetric under the corresponding exchanges, and $-1$ if they are antisymmetric \cite{Jaffe:1999ze}; then  ${\cal H}_{CM}$ can be rewritten as follows:
\be \label{oge}{\cal H}_{CM} \propto  4P_f +{4\over 3}P_s + 2P_c - {2\over 3} ~.\ee
We can see that flavor exchange, $P_f$, plays the dominant role, as it has the largest coefficient. In effect, it makes a configuration attractive whenever it is antisymmetric in flavor. In Table 1, we have included ${\cal H}_{CM}$ for each diquark configuration (in units of about $20$MeV; see \cite{Chan:1977ty,Jaffe:1999ze}), and we see that it is negative for $\Q_1$ and $\Q_2$, the diquark building blocks for mesons.

There is also an interaction under which  the attractive diquark configurations are the color-antisymmetric ones,  $\Q _1$ and $\Q _3$, as in the baryon spectrum. That interaction is the spin--independent part of OGE, also known as the color--electrostatic interaction. It is given by
\be \label{CE} {\cal H}_{CE} \propto \lambda _1 \cdot \lambda _2 = 2P_c-2/3~. \ee
As displayed in Table 1, the value of ${\cal H}_{CE}$ is $-8/3$ for the color antitriplet $\bar \3 _c$ and $4/3$ for the color sextet $\6_c$ (again in units of 20MeV); both $\Q _1$ and $\Q _3$ are attractive under this interaction and form the diquark building blocks for baryons.

We deduce that in the meson sector, the interquark forces have qualitative similarities with the spin--dependent, color-magnetic  part of OGE, and in the baryon sector the interquark forces are qualitatively similar to the spin--independent, color-electrostatic part of OGE. This distinction between the interquark forces in the meson and baryon sectors should be taken into account in the construction of any dynamical model for low--energy QCD.

 \section{Regge Trajectories of Mesons}\label{regge}

Regge trajectories are families of hadrons which have the same internal spin and isospin and the same alignment of internal spin with orbital angular momentum. They are arranged in "trajectories" of increasing orbital angular momentum $L$. The squared masses of hadrons in a trajectory are expected to increase linearly with $L$ \cite{Chew:1962eu}:
\be m^2 = a +\sigma_\alpha L ~,\ee
where $m$ is the mass of the hadron, 
and $a$ is an intercept that depends on the trajectory. The slope of the trajectory is $\sigma _\alpha$, where $\alpha$ is an index denoting the type of hadron. Here, $\alpha$ may denote  $\qq, \QQA,$ or  $\QQB$.

We list trajectories of light $\qq$, $\QQA$, and $\QQB$ mesons  in Table 4a, Table 4b, and Table 4c, respectively. We list trajectories for charmed mesons in Tables 4d and 4e, and for bottom mesons in Tables 4f and 4g. A rough approximation for the slopes $\sigma _\alpha$, which for light mesons are of order 1GeV per unit of orbital angular momentum and for heavy mesons are much higher, appears in the final row of each table. For Regge trajectories of baryons, see \cite{Wilczek:2004im,Selem:2006nd}.

\begin{table}
\small
\addcontentsline{lot}{subsection}{Table 4: Regge trajectories} 
\begin{center}{\bf Table 4a: Regge trajectories of light $\qq$}
\[ \begin{array}{|c|c|c|||c|c||c|c||c|c|}
\hline \multicolumn{9}{|c|}{\mbox{{\bf Table 4a (I): light $q\bar q$ mesons, S=1, S and L aligned}}}\\
\hline \hline
L&S&J^{PC}&[I=1]&m^2&[I=1/2]&m^2&[I=0]&m^2 \\ 
  &    &               && (GeV^2)& & (GeV^2)&& (GeV^2) \\ \hline
0&1&1^{--}&\rho(770)&0.6&K(892)&0.8&\omega(782),\phi(1020)&0.6,1.0\\
1&1&2^{++}&a_2(1320)&1.7&K_2^*(1430)&2.0&f_2(1270),f_2(1430)&1.6,2.0 \\
2&1&3^{--}&\rho_3(1690)&2.8 &K_3(1780)&3.2 &\omega _3(1670),\phi_3(1850)&2.8,3.4\\
3&1&4^{++}&a_4(2040)& 4.2&K_4^*(2045)&4.2&f_4(2050),f_4(2220)&4.2,4.9\\ 
\hline \hline
\sigma_{\qq}&&&& 1.1&&1.1& &1.2,1.2 \\ \hline
\multicolumn{9}{c}{ }
\end{array}
\]
\end{center}
\normalsize
\end{table}
\begin{table}
\small
\begin{center}
\[
\begin{array}{|c|c|c|||c|c||c|c||c|c|}
\hline \multicolumn{9}{|c|}{\mbox{{\bf Table 4a (II): light $q\bar q$ mesons, S=0}}}\\
\hline \hline
L&S&J^{PC}& [I=1]&m^2&[I=1/2]&m^2&[I=0]&m^2\\ 
 &    &               && (GeV^2)& & (GeV^2)&& (GeV^2) \\\hline
0&0&0^{-+}&\pi (135)&0.02 &K(494)&0.2& \eta(547),\eta '(958)&0.3,0.9\\
1&0&1^{+-}&b_1(1235)&1.5&K_1(1270)&1.6& h_1(1170),h_1(1380)&1.4,1.9\\
2&0&2^{-+}&\pi_2(1670)&2.8&K_2(1770)&3.1& \eta _2(1645),\eta_2(1870)&2.7,3.5\\
\hline \hline
\sigma _{\qq}&&&& 1.3&&1.5&&1.3,1.3 \\ \hline
\multicolumn{9}{c}{ }
\end{array}
\]
\end{center}
\normalsize
\end{table}
\begin{table}
\small
\begin{center}
{\bf Table 4a: Regge trajectories of light $\qq$}, continued
\[ 
\begin{array}{|c|c|c|||c|c||c|c||c|c|}
\hline \multicolumn{9}{|c|}{\mbox{{\bf Table 4a (III): light $q\bar q$ mesons, S=1, S and L antialigned}}}\\
\hline \hline
L&S&J^{PC}&  [I=1]&m^2&[I=1/2]&m^2&[I=0]&m^2 \\ 
 &    &               && (GeV^2)& & (GeV^2)&& (GeV^2) \\\hline
1&1&0^{++}&a_0(1450)&2.1&K_0^*(1430)&2.0& f_0(1370), f_0(1710)&1.9,2.9\\
2&1&1^{--}&\rho(1700)&2.9&K(1680)&2.8& \omega(1650) \hskip 1.5cm&2.7 \hskip .8cm\\
3&1&2^{++}&&&K_2^*(1980)&4.0&f_2(1950), f_2(2010)&3.8, 4.0\\
\hline \hline
\sigma _{\qq}&&&& 0.8&&1.0&&1.0, \hskip .8cm\\ \hline
\end{array}
\]
\end{center}
\normalsize
\end{table}
\begin{table}
\small
\begin{center}
\[ 
\begin{array}{|c|c|c|||c|c||c|c||c|c|}
\hline \multicolumn{9}{|c|}{\mbox{{\bf Table 4a (IV): light $q\bar q$ mesons, S=1, S and L partially aligned }}}\\ \hline \hline
L&S&J^{PC}&  [I=1]&m^2&[I=1/2]&m^2&[I=0]&m^2 \\ 
 &    &               && (GeV^2)& & (GeV^2)&& (GeV^2) \\\hline
1&1&1^{++}&a_1(1260)&1.6&K_1(1400)&2.0&f_1(1285),f_1(1420)&1.7,2.0 \\
2&1&2^{--}&&&K_2(1820)&3.3& &\\
\hline \hline
\sigma _{\qq}&&&& &&1.3&& \\ 
\hline
\end{array}
\]
\end{center}
\normalsize
\end{table}
\begin{table}
\small
\begin{center}{\bf Table 4b: Regge trajectories of light $\QQA$}
\[ \begin{array}{|c|c|c|||c|c||c|c||c|c|}
\hline \multicolumn{9}{|c|}{\mbox{{\bf light $\QQA$ mesons, S=0}}}\\
\hline \hline
L&S&J^{PC}&  [I=1]&m^2&[I=1/2]&m^2&[I=0]&m^2 \\
 &    &               && (GeV^2)& & (GeV^2)&& (GeV^2) \\ \hline
0&0&0^{++}&a_0(980)&0.8&&& f_0(600),f_0(980)&0.4,0.8\\
1&0&1^{--}&\rho(1450)&2.1&K^*(1410)&2.0& \omega(1420), \phi(1680)& 2.0, 2.8\\
2&0&2^{++}&a_2(1700)&2.9 &&& f_2(1640)\hskip 1.4cm&2.7 \hskip .7cm\\
3&0&3^{--}&\rho_3(1990)&4.0&&&& \\
\hline \hline
\sigma _{\QQA}&&&& 1.0&&&&1.2,2.0 \\ \hline
\end{array}
\]
\end{center}
\normalsize
\end{table}
\begin{table}
\small
\begin{center}{\bf Table 4c: Regge trajectories of light $\QQB$}
\[ 
\begin{array}{|c|c|c|||c|c||c|c||c|c|}
\hline \multicolumn{9}{|c|}{\mbox{{\bf light $\QQB$ mesons, S=1, S and L partially aligned}}}\\
\hline \hline
L&S&J^{PC}&[I=1]&m^2&[I=1/2]&m^2&[I=0]&m^2 \\ 
 &    &               && (GeV^2)& & (GeV^2)&& (GeV^2) \\\hline
1&1&0^{-+}&\pi (1300) &1.7 & K(1460)&2.1 & \eta (1295),\eta (1475) &1.7,2.2\\
2&1&1^{+-}& & & & &h_1(1595)\hskip 1.4cm&2.5\hskip .8cm\\
3&1&2^{-+}&\pi _2(2100) &4.4 &K_2(2250)&5.1  & &\\ \hline
\hline
\sigma _{\QQB}&&&&1.4&&1.5&&0.8 \hskip .8cm\\ 
\hline
\end{array}
\]
\end{center}
\normalsize
\end{table}
\begin{table}
\small
\begin{center}{\bf Table 4d: Regge trajectories of charmed $\qq$}
\[ 
\begin{array}{|c|c|c|||c|c||c|c||c|c|}
\hline \multicolumn{9}{|c|}{\mbox{{\bf charmed $\qq$ mesons, S=1, S and L aligned}}}\\
\hline \hline
L&S&J^{PC}&[I=1/2]&m^2&[I=0]&m^2&[I=0]&m^2 \\ 
 &    &               && (GeV^2)& & (GeV^2)&& (GeV^2) \\\hline
0&1&1^{--}&D^*&4.0 & D_s^*&4.5& J/\psi(1S) &9.6\\
1&1&2^{++}&D_2^*(2460) &6.1 &D_{s2}(2573) &6.6 &\chi_{c2}(1P) &12.6\\ \hline
\hline
\sigma ^c_{\qq}&&&&2.1&&2.1&&3\\ 
\hline
\end{array}
\]
\end{center}
\normalsize
\end{table}
\begin{table}
\small
\begin{center}{\bf Table 4e: Regge trajectories of charmed $\QQA$}
\[ 
\begin{array}{|c|c|c|||c|c||c|c||c|c|}
\hline \multicolumn{9}{|c|}{\mbox{{\bf charmed $\QQA$ mesons, S=0}}}\\
\hline \hline
L&S&J^{PC}&[I=1/2]&m^2&[I=0]&m^2&[I=0]&m^2 \\ 
 &    &               && (GeV^2)& & (GeV^2)&& (GeV^2) \\\hline
0&0&0^{++}&D_0^*(2400)&5.8 & D_{s0}^*(2317)&5.4& \chi _{c0}(1P) &11.6\\
1&0&1^{--}& & & & &\psi (2S)&13.6\\
2&0&2^{++}&& && & \chi_{c2}(2P)&15.4\\ \hline
\hline
\sigma ^c _{\QQA}&&&&&&&&1.9\\ 
\hline
\end{array}
\]
\end{center}
\normalsize
\end{table}
\begin{table}
\small
\begin{center}{\bf Table 4f: Regge trajectories of bottom $\qq$}
\[ 
\begin{array}{|c|c|c|||c|c||c|c||c|c|}
\hline \multicolumn{9}{|c|}{\mbox{{\bf bottom $\qq$ mesons, S=1, S and L aligned}}}\\
\hline \hline
L&S&J^{PC}&[I=1/2]&m^2&[I=0]&m^2&[I=0]&m^2 \\ 
 &    &               && (GeV^2)& & (GeV^2)&& (GeV^2) \\\hline
0&1&1^{--}&B^*&28.4 & B_s^*&29.3& \Upsilon (1S) &89.5\\
1&1&2^{++}&B_2^*(5747) &33.0 &B_{s2}^*(5840) &34.1 &\chi_{b2}(1P) &98.2\\ \hline
\hline
\sigma ^b_{\qq}&&&&4.6&&4.8&&8.7\\ 
\hline
\end{array}
\]
\end{center}
\normalsize
\end{table}
\begin{table}
\small
\begin{center}{\bf Table 4g: Regge trajectories of bottom $\QQA$}
\[ 
\begin{array}{|c|c|c|||c|c||c|c||c|c|}
\hline \multicolumn{9}{|c|}{\mbox{{\bf bottom $\QQA$ mesons, S=0}}}\\
\hline \hline
L&S&J^{PC}&[I=1/2]&m^2&[I=0]&m^2&[I=0]&m^2 \\ 
 &    &               && (GeV^2)& & (GeV^2)&& (GeV^2) \\\hline
0&0&0^{++}&& & && \chi _{b0}(1P) &97.2\\
1&0&1^{--}& & & & &\Upsilon (2S)&100.5\\
2&0&2^{++}&& && & \chi_{b2}(2P)&105.5\\ \hline
\hline
\sigma ^b_{\QQA}&&&&&&&&4.2\\ 
\hline
\end{array}
\]
\end{center}
\normalsize
\end{table}

\begin{quote} \end{quote}

\subsubsection*{Acknowledgements}

I am grateful to Frank Wilczek, who gave me a glimpse into his work on baryon systematics, and in response to my question "what about mesons?" encouraged me to pursue them. This work is the result. 
I am also grateful to Robert L. Jaffe, Howard Georgi, Richard Brower, Usha Mallik, Hulya Guler, Dan Pirjol, and Ayana Holloway for helpful discussions or comments.
This work was supported in part by funds provided by the U.S. Department
of Energy (DOE) under cooperative research agreement DE-FC02-94ER40818 and in part by U.S. DOE Grant number DE-FG02-91ER40685

\vskip 1cm
\appendix

\Large
\no {\bf Appendix}

\normalsize

\section{Nonet by Nonet Discussion}\label{details}
\vskip .3cm

In this appendix we provide a multiplet by multiplet discussion of the classification. 

\subsection{Light mesons}

\subsubsection*{${\bf \jpc = 0^{-+}}$}

We have two $0^{-+}$ nonets. Available assignments are an S--wave of $\qq$ and a P--wave of $\QQB$. The orbital excitation rule tells us to assign the lower-lying  nonet to the S--wave and the second nonet to the P--wave. Other $0^{-+}$ are isorons.

We took the $\eta(1475)$ to be the heavier isosinglet in the second nonet, leaving out the $\eta (1405)$. 
Our choice is due to the fact that the heavier isosinglet in any nonet should couple to kaons, and the $\eta (1475)$ couples to kaons more strongly than $\eta (1405)$ (see ``Note on $\eta (1405)$'' in \cite{Beringer:1900zz}). 

Note that the second nonet was previously taken to consist of radially excited mesons \cite{Beringer:1900zz}.

\subsubsection*{${\bf \jpc = 0^{++}}$}

 The lightest scalar nonet is $\QQA$ with $L=S=0$; an assignment of these mesons to four--quark states was suggested by Jaffe in 1976 \cite{Jaffe:1976ig}; see also \cite{Maiani:2004uc}. 

The next nonet  is the quark model's $\qq$ P--wave. The choice of isoscalar that would complete this nonet has always been ambiguous \cite{nonqq,scalarnote,Amsler:2002ey}. 
Following \cite{Amsler:2002ey}, we choose the $f_0(1710)$. The other $f_0$ mesons are isorons. 

The third (partial) nonet has masses around 2GeV, so by the orbital excitation rule it should be either a D--wave or an F--wave; the only option is a $\QQB$ D--wave.

\vskip .3cm 
\subsubsection*{${\bf \jpc =  1^{--}}$}

There are three complete or close-to-complete $1^{--}$ nonets, and two incomplete nonets which consist of only the isospin triplet (the $\rho$). Available assignments are $^3S_1$ or $^3D_1$ of $\qq$,  $^1P_1$ of $\QQA$, and $^1P_1$ or $^5P_1$ or $^5F_1$ of $\QQB$. By the orbital excitation rule, the lowest-lying  nonet, with masses less than 1GeV, is an S-wave so we assign it to $^3S_1$ of $\qq$. 

The second nonet is about .5 GeV heavier, so it is a P-wave of either $\QQA$ or $\QQB$. We assign it to $^1P_1$ of $\QQA$, though this choice is rather arbitrary -- this nonet could be a mixture of $\QQA$ and $\QQB$.

The next nonet has only the $\rho(1570)$, which appeared in the PDG for the first time in 2008. It is slightly heavy for a P-wave by the orbital excitation rule, but we still assign it to the P-wave of $\QQB$ because there is a more suitable nonet for the available D-wave assignment; since it is heavy relative to other P-waves, we choose the $^5P_1$ rather than the $^1P_1$ assignment because it is plausible that higher $S$ may mean higher mass (also see equation (\ref{oge})). 

The next nonet, which is about 1GeV higher than the lightest nonet, is a D--wave by the orbital excitation rule and we assign it to $^3D_1$ of $\qq$. Another isovector is at the mass range of F--waves, and we assign it to $\QQB$. 

Note that the second $1^{--}$ nonet was previously taken to consist of radially excited mesons \cite{Beringer:1900zz}.

\vskip .3cm
\subsubsection*{${\bf \jpc = 1^{++}}$}

There are two nonets. Available assignments are a P--wave of $\qq$ and a D-wave of $\QQB$. Using the orbital excitation rule, we assign the lighter nonet to a P--wave and the second nonet to a D--wave.

\vskip .3cm
\subsubsection*{${\bf \jpc = 1^{-+}}$}

There are no complete nonets here. However, from Table 2 we know that a $1^{-+}$ nonet should appear as $\QQB$ in a P--wave. We classify the $\pi _1 (1600)$ and $K(1630)$ as members of this nonet even though it is a bit heavy for a P--wave (we could have taken the $\pi _1(1400)$, but we opted to make the nonet consist of mesons whose masses are closer together); the $\pi _1(1400)$ is an isoron. 
Note that it has been argued  \cite{Chung:2002fz, Close:2003tv} that if the $1^{-+}$ pion  is a four--quark state, then it should be part of a large flavor multiplet, i.e. larger than a nonet. Such a multiplet has not been observed, and in our model it is not expected to be - we expect only nonets in the light flavor sector (Section \ref{afew}). 
See \cite{Beringer:1900zz,Thompson:1997bs,expion} for more on the $1^{-+}$ pions.

\vskip .3cm
\subsubsection*{${\bf \jpc = 2^{-+}}$}

There are three nonets, and there are three available assignments: a $^3P_2$ of $\QQB$, a $^1D_2$ of $\qq$, and a $^3F_2$ of $\QQB$. We assign the lightest nonet to the P-wave (even though it is a bit heavy for a P-wave), the next one to the D-wave, and the last one to the F--wave. Note that the second nonet has so far only the isovector $\pi _2 (1880)$, which in fact entered the PDG only in 2008; if it were not for its appearance, we would have assigned the lightest nonet to the D-wave based on the orbital excitation rule. 

\vskip .3cm
\subsubsection*{${\bf \jpc = 2^{--}}$}

There are two $2^-$ kaons here. We assign the lighter to a P-wave (though it's a bit heavy based on the orbital excitation rule) and the heavier to a D-wave.

\vskip .3cm
\subsubsection*{${\bf \jpc = 2^{++}}$ }

There are three almost complete nonets. The lightest and heaviest are both $\qq$, while the middle one is a $\QQA$ in a D--wave. The other two have only a single isoscalar in each; they are $\QQB$ D-waves. The $\QQB$ isoscalars and the isoscalars in the middle nonet, all D-waves, could mix. 

\vskip .3cm
\subsubsection*{${\bf \jpc = 3^{--}}$}

 There is one complete nonet, which is the D--wave of $\qq$. There are also two heavier isovectors with the same $\jpc$. Of those, the lighter one is below the baryon--antibaryon threshold, so may be $\QQA$ in an F--wave. The second is above this threshold and therefore is unlikely to have $\Q _1$ as a constituent (see decay properties, p. \pageref{moredecaydiss}); therefore, we assign it to $\QQB$ as an F--wave. 

\vskip .3cm
\subsubsection*{${\bf \jpc = 4^{++}}$}

There is one complete nonet in this sector, a $\qq$ in an F--wave. An $f_4(2300)$ should be an F-wave by the orbital excitation rule, but there are no available assignments, so it is an isoron.

\vskip .3cm
\subsubsection*{${\bf \jpc = 5^{--}}$}

The $5^{--}$ nonet could be a G-wave $\qq$ or an F--wave $\QQB$, or a $\QQA$ with even higher $L$. By the orbital excitation rule, it should be an F--wave, so we assign it to $\QQB$. However, it could be a G--wave as classified in the PDG. 
\vskip .3cm
\subsubsection*{${\bf \jpc = 6^{++}}$}

The $6^{++}$ has the mass range appropriate for an F--wave or at most a G--wave. The lowest $L$ available for this $\jpc$ is a G--wave of $\QQB$, which is our assignment. However, it could also be the H--wave as classified in the PDG.

\subsection{Charmed and bottom mesons}
\subsubsection*{${\bf \jpc = 0^{-+}}$}

There is one complete multiplet and one partial multiplet. Note that the $\jpc$ of the bottom mesons in the first multiplet have not been determined experimentally. As is standard, we assign them to S--wave of $\qq$. 

\vskip .2cm
\subsubsection*{${\bf \jpc = 0^{++}}$}

Recent suggestions (see \cite{Guler:2004pu} for reviews)
that $D_{s0}^*(2317)$ may be a tetraquark support the possibility that it completes the $\QQA$ nonet rather than the $\qq$ nonet. 

\vskip .2cm
\subsubsection*{${\bf \jpc = 1^{--}}$} 

Note that since its first appearance in the 1970's, the $\psi (2S)$ was assigned to be a radial excitation 
\cite{psi2S}. 
Until now, this assignment does not seem to have ever been questioned or challenged and is even part of the particle's name. In our paper, the $\psi (2S)$ is a diquark-antidiquark P--wave ($L=1$).


\vskip .2cm 
\subsubsection*{${\bf \jpc = 1^{++}}$}

There are two multiplets, one complete and one incomplete. Recent suggestions (see \cite{Guler:2004pu} for reviews)
that $D_{s1}^*(2460)$ may be a tetraquark support our classification to $\QQB$ rather than $\qq$. 

We classify the $X(3872)$ as a member of the $\QQB$ as well.  The $\jpc =1^{++}$ assignment for this particle is favored \cite{Abe:2005iya} but $2^{-+}$ is also possible \cite{Abulencia:2006ma}; see also \cite{CharmPDG2008,Burns:2010qq,Faccini:2012zv}. Its isospin has not been determined yet;  we listed it only under $I=0$ in the table, but its decays indicate that it must mix with $I=1$. See \cite{CharmPDG2008, Nielsen:2009uh, Close:2003sg,Godfrey:2008nc}.

Note that we include the new bottom mesons $B_1$ and $B_{s1}$; Table 14.3 of the PDG does not include them.

\vskip .2cm
\subsubsection*{${\bf \jpc = 2^{++}}$}  

There is one complete multiplet and one partial one. We include the new $B_2^*$ and $B_{s2}^*$ (which do not appear in Table 14.3 of the PDG). 

\section{"Exotic" and Outcast Mesons}

We list in Table 5 all the mesons that appear in the 2008 PDG particle listing but are not classified in Tables 14.2 and 14.3 there.\footnote{We do not include any of the mesons listed under "further states" in the PDG (those have not been confirmed).} In our model, all these mesons are no longer exotic or outcast but are part of the model and classified in our tables.

\begin{table}[h]
\renewcommand{\baselinestretch}{1.2}
\normalsize

\addcontentsline{lot}{subsection}{Table 3: "Exotic" and outcast mesons in the 2008 PDG}
\begin{center} 

{\bf Table 5: "Exotic" and Outcast Mesons in the 2008 PDG}
\end{center}
\vskip -.5cm
\small

\[ 
\begin{array}{|l|l|}
\hline 
\jpc & \\ \hline
0^{-+}&\bullet \pi (1800)  , \bullet \eta(1405) ,\eta(1760), \eta(2225)  \\
0^{++}&\bullet a_0(980) , \bullet f_0(600) , \bullet f_0(980) ,\bullet ,f_0(1500),   \\
& f_0(2020), f_0(2100),f_0(2200), f_0(2330) \\
1^{--}&\rho(1570), \rho(1900), \rho(2150), \phi(2170), \bullet \psi (4040), \\ 
& \bullet \psi (4160), Y(4260), \bullet X(4260), X(4360)\bullet \psi (4415), \\
& X(4660), \bullet \Upsilon (3S), \bullet \Upsilon (4S), \bullet \Upsilon (10860), \bullet \Upsilon (11020)\\
 1^{-+}&\bullet \pi _1 (1400), \bullet \pi _1(1600),  \\
 1^{++}&   a_1 (1640) , K_1(1650),  f_1(1510)  \\
 1^{+-}& h_1(1595) \\
 2^{-+}& \bullet \pi _2(1880),  \pi _2(2100), \\
 2^{--}&\Upsilon (1D)\\
2^{++}& f_2(1430), f_2(1565),  f_2(1640), a_2(1700), f_2(1810), \\ 
& f_2(1910),   \bullet f_2(1950), \bullet f_2(2010),f_2(2150), f_J(2220), \\
& \bullet f_2(2300), \bullet f_2(2340), \chi _{c2}(2P) \\
3^{--}& \rho_3 (1990), \rho_3 (2250) \\
4^{++}&f_4(2300)\\
\hline
\end{array} 
\hskip .2cm
\begin{array}{|l|l|}
\hline
J^P & \\ \hline
0^-& K(1830)\\
0^+&\kappa (800) , K_0^*(1950),\\ 
1^+ &K_1(1650), \bullet B_1(5721)^0, \bullet B_{s1}(5830)^0\\
2^-& K_2(1580), K_2(2250)\\
2^+&K_2^*(1980), B_2^*(5747)^0, B_{s2}^*(5840)^0 \\
3^+&K_3(2320)\\
4^-&K_4(2500)\\
5^-& K_5^*(2380)\\
?^?&K(1630), K(3100), D^*(2640),  \\ 
& Y(3940),  B^*_J(5732),  B^*_{sJ}(5850), \\
& h_c(1P)\\ 
\hline \multicolumn{2}{c}{ }\\
\multicolumn{2}{c}{ }\\
\multicolumn{2}{c}{ }\\
\multicolumn{2}{c}{ }\\
\multicolumn{2}{c}{ }\\
\end{array}
\]
\end{table}

\end{document}